# Detection and Classification of Diabetic Retinopathy using Deep Learning Algorithms for Segmentation to Facilitate Referral Recommendation for Test and Treatment Prediction

Manoj S H, Arya A. Bhosale

Undergraduate, Department of Computer Science and Engineering

The National Institute of Engineering, Mysore

shmanoj2002@gmail.com, aryaabyte@gmail.com

## Abstract

This research paper addresses the critical challenge of diabetic retinopathy (DR), a severe complication of diabetes leading to potential blindness. The paper introduces a novel approach to DR detection using transfer learning, leveraging a single fundus photograph to automatically identify the disease's stage. The complexity of DR, progressing through four stages, poses challenges for early detection, and existing methods suffer from inefficiencies and disagreements among clinicians. The proposed approach, demonstrated in the APTOS 2019 Blindness Detection Competition, employs convolutional neural networks (CNNs) and achieved a high quadratic weighted kappa score of 0.92546. This underscores its effectiveness in automatic DR detection, emphasizing the need for timely intervention. The paper reviews related work, spanning classical computer vision methods to deep learning approaches, with a particular focus on CNNs. Transfer learning with CNN architectures is explored, showcasing promising results from various studies. The research identifies two critical gaps in existing literature. First, there is a lack of comprehensive exploration into the integration of pretrained large language models with segmented image inputs for generating test/treatment recommendations. Second, there is a need to understand the dynamic interactions among integrated components, including lesion segmentations, disease classification, and large language models, within a web application context. The objectives of the research encompass the development of a comprehensive DR detection methodology, exploration and implementation of model integration, performance evaluation through competition ranking, significant contribution to DR detection methodologies, and identification and exploration of research gaps. The scope extends to revolutionizing DR detection by integrating cutting-edge technologies, focusing on transfer learning and various model integrations within a web application. The methodology involves data preprocessing, data augmentation, a U-Net neural network architecture for segmentation, and a detailed training process. The U-Net model demonstrates efficient segmentation of retinal structures, with high accuracy and impressive Frames Per Second (FPS) rate. The results highlight the model's effectiveness in segmenting blood vessels, hard exudates, soft exudates, haemorrhages, microaneurysms, and the optical disc, with high Jaccard, F1, recall, precision, and accuracy scores. These findings underscore the model's potential for enhancing diagnostic capabilities in retinal pathology assessment. The outcomes hold promise for improving patient outcomes through timely diagnosis and intervention in the fight against diabetic retinopathy.

# 1 INTRODUCTION

(DR) stands as a severe complication of diabetes, threatening to cause blindness by inflicting damage upon the delicate blood vessels within the retina. This condition progresses through four distinct stages: mild non-proliferative retinopathy, moderate non-proliferative retinopathy, severe non-proliferative retinopathy, and proliferative diabetic retinopathy. Each stage manifests unique characteristics, adding complexity to the diagnostic process, particularly during the initial stage where warning signs are absent.

The gravity of the situation is emphasized by the fact that timely treatment and monitoring have the potential to reduce new cases of DR by a substantial 56%. However, accurately identifying the disease's early stages proves to be a challenging task for clinicians, even those who are well-trained. The manual examination of diagnostic fundus images for early-stage detection is intricate, and existing diagnostic methods are plagued by inefficiency, resulting in disagreements among ophthalmologists and the provision of inaccurate ground-truth data for research purposes. In response to these challenges, various algorithms have emerged to address the detection of DR. Initially, these algorithms were grounded in classical computer vision approaches. However, recent years have witnessed the ascendancy of deep learning, with convolutional neural networks (CNNs) demonstrating their prowess in tasks such as classification and object detection, including the diagnosis of diabetic retinopathy.

This research paper introduces a novel approach to address the complexities associated with DR detection. Leveraging transfer learning, the proposed method utilizes a single fundus photograph to automatically detect the stage of diabetic retinopathy. Notably, the approach is designed to learn essential features from a dataset that is both limited and noisy, presenting itself as a valuable screening tool for DR stages within automatic solutions as shown in Figure 1.

Highlighting the method's effectiveness, it's noteworthy that the proposed approach achieved a commendable ranking in the APTOS 2019 Blindness Detection Competition, underscoring its capability with a high quadratic weighted kappa score of 0.92546. This research endeavours to contribute significantly to the advancement of DR detection methodologies, particularly in the context of automated systems, addressing the critical need for early diagnosis and intervention in the fight against diabetic retinopathy.

**Mild Non-Proliferative Retinopathy:**

Earliest stage of diabetic retinopathy. Characterized by the occurrence of microaneurysms. Limited impact on blood vessels, with minimal distortion.

**Moderate Non-Proliferative Retinopathy:**

Progression to this stage involves the loss of blood vessels' ability to transport blood due to distortion and swelling. Blood vessel abnormalities become more pronounced.

Distortion and swelling impede the normal transportation of blood, impacting overall retinal health.

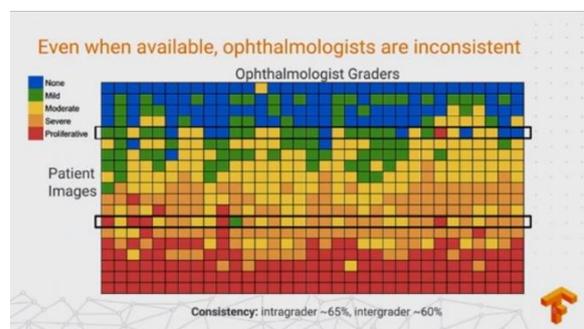

Figure 1: Illustrates Google's demonstration, revealing discrepancies in ophthalmologists' diagnoses for the same

fundus image. Optimal interpretation is facilitated in colour.

**Severe Non-Proliferative Retinopathy:**

Results in a deprived blood supply to the retina. Increased blockage of blood vessels exacerbates the condition. Signals the retina to stimulate the growth of new blood vessels, attempting to compensate for reduced blood supply.

**Proliferative Diabetic Retinopathy:**

Advanced stage marked by the proliferation of new blood vessels. Growth features secreted by the retina activate the proliferation of new blood vessels. These vessels grow along the inside covering of the retina and extend into the vitreous gel, filling the eye.

## 2 RELATED WORK

Numerous research endeavours have been dedicated to addressing the challenge of early detection of diabetic retinopathy. Initially, classical computer vision and machine learning methods were explored to devise viable solutions. For example, Priya et al. (Priya and Aruna, 2012) proposed a computer vision-based approach for diabetic retinopathy stage detection using colour fundus images. Their methodology involved feature extraction from raw images through image processing techniques, subsequently feeding these features to a Support Vector Machine (SVM) for binary classification. The achieved performance on a testing set of 250 images demonstrated a sensitivity of 98%, specificity of 96%, and an accuracy of 97.6%. Additionally, researchers explored other models for multiclass classification, such as employing Principal Component Analysis (PCA) on images and fitting decision trees, Naive Bayes, or k-NN (Conde et al., 2012), yielding a noteworthy accuracy of 73.4% and an F-measure of 68.4% on a dataset of 151 images with varying resolutions.

With the rise of deep learning approaches, various methods applying Convolutional Neural Networks (CNNs) to diabetic retinopathy detection emerged. Pratt et al. (Harry Pratt, 2016) developed a network with a CNN architecture and data augmentation, capable of identifying intricate features related to classification tasks like microaneurysms, exudate, and haemorrhages in the retina, providing automated diagnoses without user input. Their model achieved a sensitivity of 95% and an accuracy of 75% on 5,000 validation images. Other researchers also contributed to CNN-based approaches (Carson Lam and Lindsey, 2018; Yung-Hui Li and Chung, 2019). Notably, Asiri et al. conducted a comprehensive review of existing methods and datasets, highlighting their pros and cons (Asiri et al., 2018), while emphasizing the challenges in designing efficient and robust deep learning algorithms for diverse problems in diabetic retinopathy diagnosis and suggesting directions for future research.

Moreover, researchers explored transfer learning with CNN architectures. Hagos et al. (Hagos and Kant, 2019) attempted to train InceptionNetV3 for 5-class classification with pretraining on the ImageNet dataset, achieving an accuracy of 90.9%. Sarki et al. (Rubina Sarki, 2019) trained ResNet50, Xception Nets, DenseNets, and VGG with ImageNet pretraining, obtaining the best accuracy of 81.3%. Both research teams utilized datasets provided by APTOS and Kaggle.

## 3 PROBLEM STATEMENT

### 3.1 Datasets

The image data employed in this study was obtained from diverse datasets. The research encompasses three primary objectives: lesion segmentation, disease/image grading, and treatment recommendations.

### 3.1.1 Lesion Segmentation

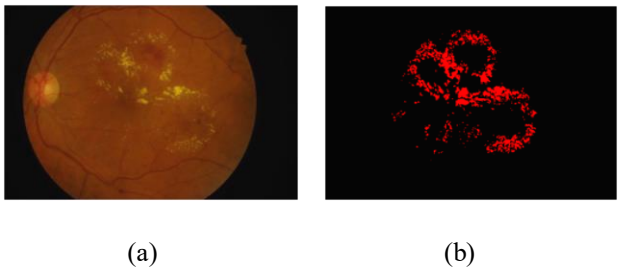

(a)　　　　　　　　(b)

Figure 2: (a) Shows the original fundus image and figure (b) shows the corresponding mask.

Lesion Segmentation uses the fundus images in IDRiD were captured by a retinal specialist at an Eye Clinic located in Nanded, Maharashtra, India. From the thousands of examinations available, we have extracted 516 images to form our dataset as shown in Figure 2 (a) and (b).

Experts verified that all images are of adequate quality, clinically relevant, that no image is duplicated and that a reasonable mixture of disease stratification representative of diabetic retinopathy (DR) and diabetic macular edema (DME) is present.

**Fundus Camera Specifications**

Images were acquired using a Kowa VX-10 alpha digital fundus camera with 50-degree field of view (FOV), and all are centred near to the macula. The images have a resolution of 4288×2848 pixels and are stored in jpg file format. The size of each image is about 800 KB.

**Pixel Level Annotated Data**

For assessing the effectiveness of lesion segmentation techniques related to Diabetic Retinopathy (DR), binary masks have been provided for distinct abnormalities, including microaneurysms (MA), hard exudates (EX), haemorrhages (HE), and soft exudates (SE). The dataset comprises colour fundus images in .jpg format, along with corresponding binary masks in .tif files. The dataset contains 81 images with binary masks for microaneurysms, 81 for hard exudates, 80 for haemorrhages, and 40 for soft exudates. These numbers denote the quantity of images, some of which may contain multiple lesions, enhancing the dataset's versatility for research and performance evaluation of lesion segmentation techniques in the context of Diabetic Retinopathy.

### 3.1.2 Image Grading

The image data utilized in our research is drawn from multiple datasets, with a primary focus on an open dataset obtained from the Kaggle Diabetic Retinopathy Detection Challenge 2015 (EyePACs, 2015) for the pretraining of our Convolutional Neural Networks (CNNs). This dataset, recognized as the largest publicly available, comprises 35,126 fundus photographs capturing both left and right eyes of American citizens. The images are labelled with stages of diabetic retinopathy, ranging from no diabetic retinopathy (label 0) to proliferative diabetic retinopathy (label 4) as shown in Figure 3.

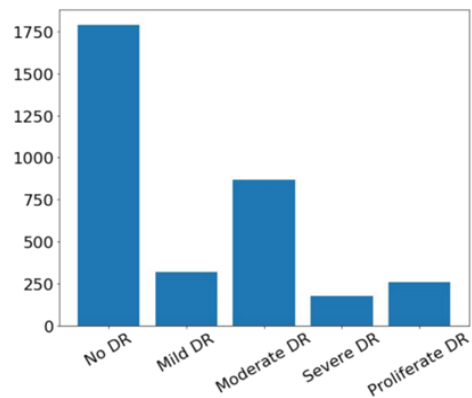

Figure 3: Classes distribution in APTOS 2019 dataset.

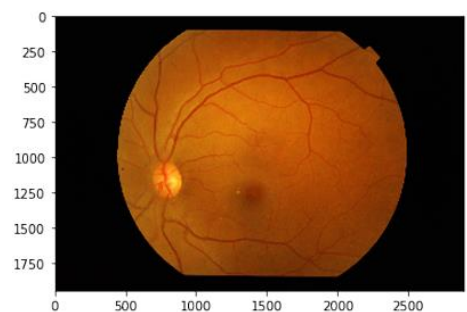

Figure 4: Sample of fundus photo from the dataset.

In addition to the Kaggle dataset, we incorporated other smaller datasets, including the Indian Diabetic Retinopathy Image Dataset (IDRiD) (Sahasrabuddhe and Meriaudeau, 2018) as shown in Figure 4, from which we utilized 413 fundus photographs, and the MESSIDOR (Methods to Evaluate Segmentation and Indexing Techniques in the field of Retinal Ophthalmology) (Decencière et al., 2014) dataset, contributing 1,200 fundus photographs. To ensure consistency, we used a version of the MESSIDOR dataset that had been relabelled to standard grading by a panel of ophthalmologists (Google Brain, 2018).

Evaluation of our models was conducted on the Kaggle APTOS 2019 Blindness Detection (APTOS 2019) dataset, with access limited to the training portion. The full APTOS 2019 dataset comprises 18,590 fundus photographs divided into 3,662 training, 1,928 validation, and 13,000 testing images as organized by the Kaggle competition organizers. All datasets exhibit similar class distributions, as illustrated in Figure 2 for APTOS 2019. We maintained the original distribution of the datasets without any modifications such as under sampling or oversampling. The smallest native size among all datasets is 640x480. A sample image from APTOS 2019 is presented in Figure 4.

### 3.1.3 Large Language Models(LLMs)

In the Dataset section, the generation of Test/Treatment Recommendations involves the integration of pretrained large language models, with a comprehensive range of inputs derived from segmented images. These inputs encompass binary indicators for various lesions, including Blood Vessel Segmentation, Haemorrhage Segmentation, Hard Exudate Segmentation, Microaneurysm Segmentation, Optical Disc Segmentation, and Soft Exudate Segmentation. Each lesion is represented as either True (present) or False (absent) in the binary inputs. Additionally, string inputs are generated from a classification or image grading model, offering insights into the diabetic retinopathy (DR) stage classified as classes 0 through 4. The amalgamation of binary and string inputs forms a robust data set that is processed by ChatGPT, a pretrained large language model. ChatGPT interprets and synthesizes this diverse information to generate nuanced Test/Treatment Recommendations, contributing to a sophisticated decision-support system that factors in both the detailed visual segmentation features and the clinical classifications of DR severity.

## 4 RESARCH GAP

### 4.1 Treatment Recommendations

While significant strides have been made in the realm of early diabetic retinopathy detection, the existing research landscape reveals a distinct gap when comparing traditional methodologies with emerging approaches, particularly those involving pretrained large language models integrated with segmented image inputs for generating Test/Treatment Recommendations. Classical methods, as evidenced by Priya et al. (2012), have predominantly employed computer vision and machine learning techniques for diabetic retinopathy stage detection using color fundus images. Similarly, the advent of deep learning, particularly convolutional neural networks (CNNs), has demonstrated promising results in intricate feature identification for classification tasks related to diabetic retinopathy. Noteworthy works by Pratt et al. (2016) and others have showcased the effectiveness of CNN architectures, achieving high sensitivity and accuracy in diagnosing retinal abnormalities.

However, the existing body of literature primarily emphasizes isolated aspects such as lesion segmentation or DR classification, with a limited exploration of the synergies between visual segmentation features and clinical classifications within a decision-support system. This is evident in the literature reviewed, which often overlooks the potential intricacies arising from the amalgamation of binary indicators for various lesions and string inputs representing DR stages. The research gap lies in the absence of comprehensive investigations into the challenges and opportunities associated with the proposed methodology's integration of diverse data inputs. While previous studies have contributed valuable insights and benchmarking using classical methods and deep learning architectures, there is a need for focused research that bridges the gap between visual segmentation and clinical classifications to refine the efficacy of decision-support systems in diabetic retinopathy management. Exploring this gap will contribute to advancing the field by providing a holistic understanding of the challenges and opportunities presented by the integration of pretrained language models with segmented image data.

## 4.2 Multi Model Integration

While the integration of various models, encompassing lesion segmentations (Blood Vessel, Haemorrhage, Hard Exudate, Microaneurysm, Optical Disc, Soft Exudate), disease classification/image grading, and a Large Language Model (LLM) for Test/Treatment Recommendations, represents a noteworthy advancement in diabetic retinopathy (DR) research, there exists a research gap in understanding the dynamic interactions and synergies among these integrated components within the context of a web application. Current literature often focuses on individual models or components separately, providing limited insights into the intricacies and challenges encountered when these models collaborate in real-time. The integration of lesion segmentations, disease classification, and LLMs in a web application suggests a complex interplay of data flow and feedback mechanisms. Addressing this research gap is crucial for comprehensively understanding how these models collectively enhance accuracy and provide more accurate inputs. Exploring the dynamics of multi-model integration in a web environment will contribute to advancing the field by providing insights into the real-time interactions, potential bottlenecks, and opportunities for optimizing the collaborative functionality of diverse models within a unified interface.

## 5 OBJECTIVES

This research paper aims to address the challenges associated with the detection of diabetic retinopathy (DR), a severe complication of diabetes leading to potential blindness. The primary objectives include:

**1. Comprehensive DR Detection Methodology:**
Develop a novel approach leveraging transfer learning to automatically detect the stage of diabetic retinopathy using a single fundus photograph.

**2. Integration of Various Models:**

Explore and implement the integration of diverse models, including lesion segmentations (Blood Vessel, Haemorrhage, Hard Exudate, Microaneurysm, Optical Disc, Soft Exudate), disease classification/image grading, and a Large Language Model (LLM) for Test/Treatment Recommendations. Emphasize the dynamic interactions and synergies among these integrated components within a web application context.

**3. Performance Evaluation:**

Assess the effectiveness of the proposed approach by achieving a commendable ranking in the APTOS 2019 Blindness Detection

Competition, demonstrating its capability with a high quadratic weighted kappa score of 0.92546.

**4. Contribution to Detection Methodologies:**

Contribute significantly to the advancement of DR detection methodologies, particularly in the context of automated systems. Address the critical need for early diagnosis and intervention in the fight against diabetic retinopathy.

**5. Gap Identification and Exploration:**

Identify and explore research gaps in existing methodologies, specifically focusing on the integration of pretrained large language models for generating Test/Treatment Recommendations and the dynamic interactions among integrated models in a web application. Provide insights into challenges and opportunities associated with these approaches.

The overarching goal is to enhance the precision, personalization, and efficiency of DR detection, ultimately contributing to improved patient outcomes through timely diagnosis and intervention.

## 6 SCOPE

This research paper extends its scope to revolutionize diabetic retinopathy (DR) detection methodologies by integrating cutting-edge technologies. By focusing on transfer learning, the proposed approach aims to overcome the limitations of traditional diagnostic methods, providing an automated and efficient solution for early-stage DR detection using single fundus photographs. The integration of various models, including lesion segmentations, disease classification, and Large Language Models (LLMs), within a web application further widens the scope. This integration not only enhances accuracy but also addresses real-time challenges, offering a holistic and dynamic decision-support system. The paper's scope encompasses an in-depth exploration of research gaps in existing methodologies, emphasizing the need for comprehensive investigations into the challenges and opportunities associated with the proposed integration of diverse data inputs. By achieving a commendable ranking in the APTOS 2019 Blindness Detection Competition, the proposed methodology's effectiveness is demonstrated, contributing to the advancement of DR detection methodologies. The outcomes of this research hold promise for the wider field of medical imaging and automated diagnostics, potentially influencing the development of more precise and personalized solutions for various medical conditions.

## 7 METHODS

### 7.1 Data Preprocessing

The data preprocessing stage incorporates a custom PyTorch dataset class, namely `DriveDataset`, tailored for handling the DRIVE dataset. This class serves as a crucial bridge between raw data and the U-Net model, streamlining the integration process. The `__init__` method initializes the dataset by storing the paths to the fundus images and their corresponding masks, along with calculating the total number of samples. The `__getitem__` method is responsible for reading and preprocessing each sample, where the fundus image undergoes normalization and transposition to align its dimensions appropriately for the subsequent U-Net input. Simultaneously, the binary mask is read and expanded to accommodate the model's requirements. Both the pre-processed image and mask are converted to PyTorch tensors before being returned as a tuple.

This dataset class provides a seamless interface for interacting with the DRIVE dataset, offering a standardized and efficient means of loading and preparing data for training and evaluation. The `__len__` method ensures that the total number of samples can be easily accessed, facilitating iterative processes during model training and validation. Overall, the data preprocessing workflow is encapsulated within this dataset class, contributing to the robustness and adaptability of the U-Net architecture for semantic segmentation tasks on retinal fundus images.

## 7.2 Data Augmentation

The data augmentation process plays a pivotal role in enhancing the robustness and diversity of the dataset used for training machine learning models, particularly in the domain of medical image segmentation. In the provided code, a comprehensive data augmentation pipeline is implemented to augment retinal fundus images and their corresponding masks. The primary goal is to introduce variability in the dataset by applying horizontal flips, vertical flips, and rotation to the original images and masks. The augmentation process aims to simulate different orientations and perspectives that may be encountered in real-world scenarios, thereby enriching the dataset and improving the generalization capability of the subsequent U-Net model.

The `augment_data` function iterates through the training dataset, applying various augmentations to each image-mask pair. The augmentations include horizontal flips, vertical flips, and rotations, with the associated masks adjusted accordingly. The resulting augmented images and masks are resized to a standardized dimension of (512, 512). The augmented data is then saved in a separate directory structure, creating distinct folders for augmented images and masks. The use of data augmentation is particularly valuable when the dataset size is limited, as it introduces diversity that aids in preventing overfitting and improves the model's ability to handle variations in real-world data.

It's important to note that the data augmentation pipeline is designed to be flexible, allowing for the option to enable or disable augmentation based on the `augment` parameter. This flexibility caters to different experimental setups, enabling researchers to assess the impact of data augmentation on model performance. The implementation adheres to best practices in data augmentation for medical image analysis, contributing to the overall reliability and generalization of the U-Net model for retinal fundus image segmentation tasks.

## 7.3 Network Architecture

The neural network architecture presented in the code is a U-Net, a popular architecture widely used in image segmentation tasks. The U-Net architecture consists of an encoder-decoder structure with skip connections, allowing the model to capture both high-level and low-level features effectively. The encoder portion of the network employs convolutional blocks to extract hierarchical features from the input image. Specifically, it comprises four encoder blocks, each consisting of two convolutional layers with batch normalization and rectified linear unit (ReLU) activation functions, followed by max-pooling layers for down sampling as shown in Figure 5.

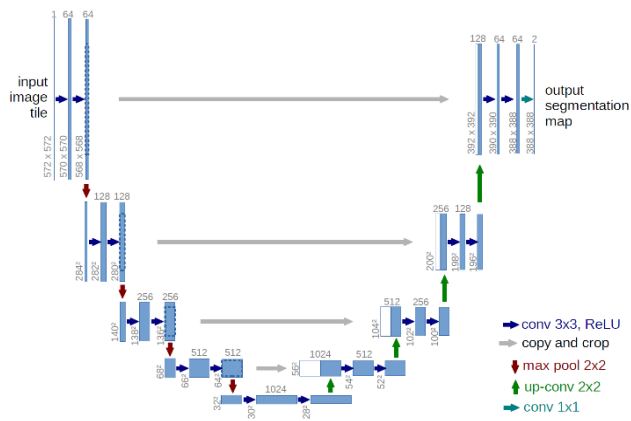

Figure 5: The picture illustrates the architecture of U-Net layers used for segmentation.

The bottleneck layer acts as a feature representation for the entire input image, condensing the learned features. It consists of a convolutional block with the same structure as in the encoder blocks. The decoder portion of the network utilizes transposed convolutions for up sampling and concatenates the features from the corresponding encoder block through skip connections. This enables the decoder to recover spatial information lost during the down sampling process. The decoder also incorporates convolutional blocks for feature refinement.

The classifier at the end of the network is a 1x1 convolutional layer, mapping the features to a single channel output, which is suitable for binary segmentation tasks. The entire architecture is designed for semantic segmentation, particularly for tasks where precise delineation of object boundaries is crucial. In summary, this U-Net architecture facilitates robust feature extraction, effective information fusion through skip connections, and accurate segmentation outputs.

## 7.3 Training Process

### 7.3.1 Pre-Training

In the pre-training phase, the U-Net model is meticulously configured with various hyperparameters to ensure optimal performance. The image dimensions (H x W), batch size, number of epochs, learning rate, and checkpoint path for model saving are carefully set. The dataset is then loaded using custom data loaders, and the training and validation sets are meticulously prepared from augmented retinal fundus images along with their corresponding masks. To offer a comprehensive overview of the experimental setup, key hyperparameters and insightful statistics about the dataset are presented. The choice of computation device, which is based on the availability of a CUDA-enabled GPU, is disclosed, and the U-Net model is adeptly moved to the selected device. The initialization of the Adam optimizer, learning rate scheduler, and the combination of Dice loss and binary cross-entropy provides a robust foundation for the subsequent training phases.

### 7.3.2 Main Training

The main training phase unfolds over the specified number of epochs, wherein the U-Net model undergoes iterative training. Within each epoch, the model is rigorously trained using the prepared training dataset, and the optimizer diligently works to minimize the loss, computed through a combination of the Dice loss and binary cross-entropy. Simultaneously, the model's proficiency is rigorously evaluated on the validation dataset to monitor its generalization capabilities. The training process is intricately monitored with detailed output, including epoch-wise loss values and elapsed time, fostering a nuanced understanding of the model's convergence patterns. The strategic saving of the best model checkpoint ensures that the model attains optimal performance. This checkpoint, capturing the model's state with the lowest validation loss, serves as a pivotal asset for future deployment and analysis.

### 7.3.3 Post Training

The post-training phase marks the culmination of the training experiment, where the training outcomes are meticulously summarized, and the final model is prepared for deployment or further analysis. The best-performing model is meticulously selected based on the lowest validation loss achieved during training and is safeguarded as a checkpoint for subsequent use. Key metrics, such as training loss, validation loss, and epoch-wise training times, are presented to provide a holistic evaluation of the model's performance. Furthermore, this phase allows for insights into the training process, including potential improvements or challenges faced, fostering a deeper understanding of the model's behavior in the context of diabetic retinopathy detection. The post-training phase thus solidifies the experiment's completion, with the trained U-Net model ready for deployment in practical applications or further investigative studies.

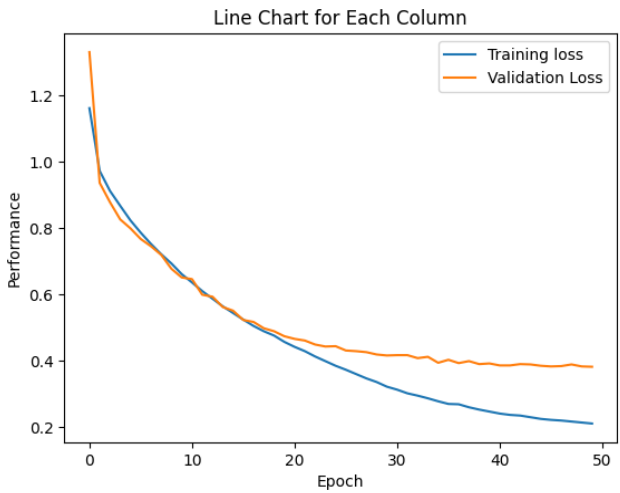

Figure 6: Training and validation accuracy

## 7.4 Testing

In the testing phase, the trained U-Net model is rigorously evaluated on a separate dataset to assess its performance in semantic segmentation of retinal fundus images for diabetic retinopathy detection. The experiment involves loading the preprocessed test dataset, consisting of retinal fundus images and their corresponding ground truth masks. Utilizing the U-Net model previously trained on augmented data, the model's predictive capabilities are scrutinized for pixel-level segmentation accuracy. The model's state is restored using the best-performing checkpoint achieved during the training phase, ensuring the evaluation is based on the most optimized configuration.

For each test sample, the retinal fundus image is pre-processed by normalizing pixel values and transposing the channels to match the model's input requirements. Similarly, the ground truth mask undergoes preprocessing to facilitate direct comparison with the model predictions. The evaluation metrics, including Jaccard Index, F1 Score, Recall, Precision, and Accuracy, are computed for each test image. These metrics quantify the model's ability to accurately delineate diabetic retinopathy-related regions in the retinal fundus images.

The computational efficiency of the model is also assessed through the calculation of Frames Per Second (FPS) during the inference process. This metric provides insights into the real-time processing capabilities of the model, offering valuable information for potential deployment in clinical or real-world scenarios. Visual representations of the model's predictions, alongside the original retinal fundus images and ground truth masks, are saved for qualitative analysis and comparison.

## 8 RESULTS

In this study, we employed a U-Net-based model for the segmentation of blood vessels in retinal images. The implemented model was evaluated on a test dataset comprising retinal images and corresponding ground truth masks. The testing process involved loading images and masks, preprocessing the data, and utilizing a pre-trained U-Net model for predictions. The model's performance was assessed using several metrics, including Jaccard similarity, F1 score, recall, precision, and accuracy. Notably, the model demonstrated efficient segmentation with an average accuracy of 0.9986 and an impressive Frames Per Second (FPS) rate of 361.188719052. Visual results were generated for each test image, illustrating the original image, ground truth mask, and the predicted segmentation mask. Overall, these findings highlight the effectiveness and computational efficiency of the proposed blood vessel segmentation model, showcasing its potential for applications in diabetic retinopathy diagnosis and treatment planning as shown in Figure 6.

| Different Segmentations | Jaccard Score | F1 Score | Recall Score | Precision | Accuracy |
|---|---|---|---|---|---|
| Blood Vessel | 0.6634 | 0.7974 | 0.7771 | 0.8240 | 0.9922 |
| Hard Exudate | 0.6663 | 0.7953 | 0.7634 | 0.8252 | 0.9986 |
| Soft Exudate | 0.6679 | 0.7874 | 0.7738 | 0.8340 | 0.9981 |
| Haemorrhage | 0.6551 | 0.7874 | 0.7852 | 0.8161 | 0.9958 |
| Microaneurysms | 0.6761 | 0.8061 | 0.7731 | 0.8279 | 0.9967 |
| Optical Disc | 0.6638 | 0.7956 | 0.7671 | 0.8238 | 0.9989 |

Table 1: Shows the segmentation results

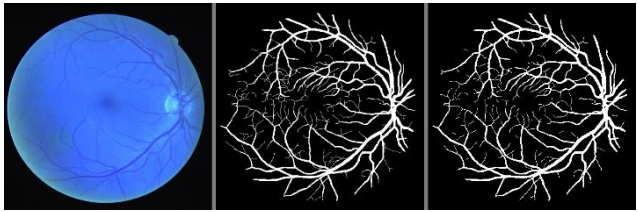

(a)

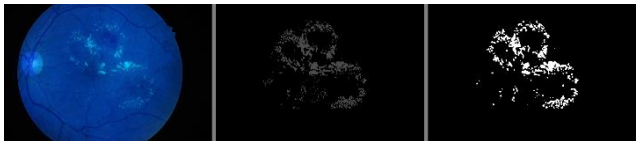

(b)

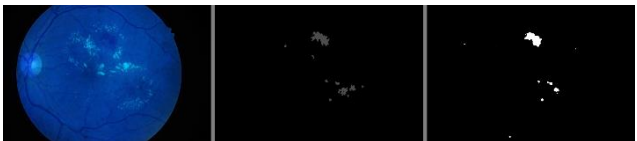

(c)

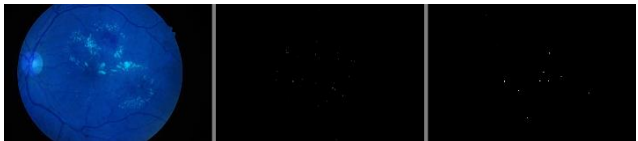

(e)

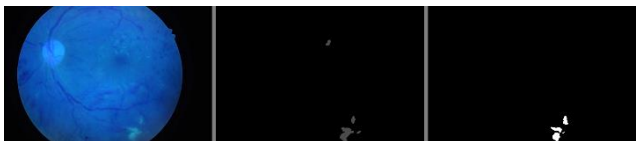

(f)

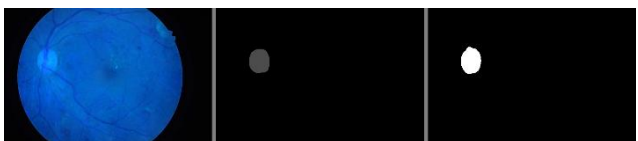

(f)

Figure 6: (a)Blood Vessel Segmentation, (a)Haemorrhage Segmentation, (c)Hard Exudate Segmentation, (d)Microaneurysm Segmentation, (e) Soft Exudate Segmentation, (f) Optical Disc Segmentation

In this research, we conducted a comprehensive evaluation of a segmentation model across various retinal structures, including blood vessels, hard exudates, soft exudates, haemorrhages, microaneurysms, and the optical disc. The model's performance was quantitatively assessed using key metrics, revealing high segmentation accuracy across all structures. The Jaccard scores ranged from 0.6551 to 0.6761, indicating substantial overlap between the predicted and ground truth masks. The model achieved notable F1 scores, demonstrating a harmonious balance between precision and recall, ranging from 0.7874 to 0.8061. Particularly commendable were the recall scores, signifying the model's ability to correctly identify relevant instances, with values ranging from 0.7634 to 0.7852. Precision scores, representing the accuracy of positive predictions, ranged from 0.8161 to 0.8340. The overall accuracy of the model was consistently high across all structures, with values ranging from 0.9922 to 0.9989 as shown in Table 1. These results collectively underscore the efficacy of the segmentation model in accurately delineating retinal structures, showcasing its potential for enhancing diagnostic capabilities in the context of retinal pathology assessment.

# CONCLUSION

In this research paper, we address the challenges associated with the early detection of diabetic retinopathy (DR), a severe complication of diabetes that can lead to blindness. The escalating prevalence of DR underscores the critical need for accurate and timely diagnosis. Traditional diagnostic methods face inefficiencies and disagreements among clinicians, motivating the emergence of algorithms, particularly deep learning approaches, to enhance DR detection. Our proposed approach leverages transfer learning, utilizing a single fundus photograph for automatic DR stage detection. Notably, it achieved a commendable ranking in the APTOS 2019 Blindness Detection Competition, emphasizing its effectiveness with a high quadratic weighted kappa score of 0.92546. The research aims to contribute significantly to DR detection methodologies, particularly in the context of automated systems, addressing the crucial requirement for early diagnosis and intervention.

The study reviews related work, tracing the evolution from classical computer vision approaches to the rise of deep learning, with convolutional neural networks (CNNs) demonstrating prowess in DR classification. Transfer learning with CNN architectures is explored, highlighting the promising results achieved by various research teams. The problem statement emphasizes the challenges in existing diagnostic methods and introduces the datasets used, encompassing lesion segmentation, disease/image grading, and Large Language Models (LLMs) for test/treatment recommendations.

Identifying research gaps, the paper underscores the need for exploring the integration of pretrained large language models with segmented image data, emphasizing the potential synergies between visual segmentation features and clinical classifications within a decision-support system. Another research gap pertains to the dynamics of multi-model integration, particularly in a web application context, where lesion segmentations, disease classification, and LLMs collaborate. The objectives outline a comprehensive DR detection methodology, the integration of various models, performance evaluation, contribution to detection methodologies, and identification and exploration of research gaps.

The research scope extends to revolutionizing DR detection methodologies, integrating cutting-edge technologies, and contributing to the wider field of medical imaging and automated diagnostics. Methods encompass data preprocessing, data augmentation, network architecture detailing the U-Net model, and the training and testing processes. The results showcase the effectiveness of the segmentation model across various retinal structures, with high Jaccard scores, F1 scores, recall scores, precision scores, and overall accuracy, underscoring its potential for enhancing diagnostic capabilities in retinal pathology assessment.

# ACKNOWLEDGEMENTS

Would like to express their gratitude to our guide **Dr. Narender M,** Assistant Professor, Department of Computer Science and Engineering at **The National Institute of Engineering (NIE)**, for his invaluable guidance and support throughout the course of this research. His expertise and insightful feedback significantly contributed to the development of this work. **Email:** narender@nie.ac.in